\documentclass[twocolumn,showpacs,preprintnumbers,amsmath,amssymb,nofootinbib]{revtex4}

\usepackage{amsmath}
\usepackage{graphicx}
\usepackage{calc}
\usepackage{bm}
\usepackage{float}

\begin{document}

\title{Manifestation of strong correlations in transport in ultra-clean SiGe/Si/SiGe quantum wells}
\author{A.~A. Shashkin}
\affiliation{Institute of Solid State Physics, Chernogolovka, Moscow District 142432, Russia}
\author{M.~Yu.\ Melnikov}
\affiliation{Institute of Solid State Physics, Chernogolovka, Moscow District 142432, Russia}
\author{V.~T. Dolgopolov}
\affiliation{Institute of Solid State Physics, Chernogolovka, Moscow District 142432, Russia}
\author{M.~M. Radonji\'c}
\affiliation{Institute of Physics Belgrade, University of Belgrade, Pregrevica 118, 11080 Belgrade, Serbia}
\author{V. Dobrosavljevi\'c}
\affiliation{Department of Physics and National High Magnetic Field Laboratory, Florida State University, Tallahassee, Florida 32306, USA}
\author{S.-H. Huang and C.~W. Liu}
\affiliation{Department of Electrical Engineering and Graduate Institute of Electronics Engineering, National Taiwan University, Taipei 106, Taiwan\\ and National Nano Device Laboratories, Hsinchu 300, Taiwan}
\author{Amy Y.~X. Zhu}
\affiliation{Physics Department, Northeastern University, Boston, Massachusetts 02115, USA}
\author{S.~V. Kravchenko}
\affiliation{Physics Department, Northeastern University, Boston, Massachusetts 02115, USA}

\begin{abstract}
We observe that in a strongly interacting two-dimensional electron system in ultra-clean SiGe/Si/SiGe quantum wells, the resistivity on the metallic side near the metal-insulator transition increases with decreasing temperature, reaches a maximum at some temperature, and then decreases by more than one order of magnitude. We scale the resistivity data in line with expectations for the transport of strongly correlated Fermi systems and find a nearly perfect agreement with theory over a wide range of electron densities.
\end{abstract}
\pacs{71.10.-w, 71.27.+a, 71.30.+h}
\maketitle

Much interest has recently been directed toward the behavior of low-disorder, strongly interacting electrons in two dimensions (2D), for which the interaction parameter $r_{\text s}=1/(\pi n_{\text{s}})^{1/2}a_{\text{B}}$ greatly exceeds unity (here $n_{\text{s}}$ is the areal density of electrons and $a_{\text{B}}$ is the effective Bohr radius in semiconductor). These systems are characterized by the strong metallic temperature dependence of the resistivity at sub-kelvin temperatures \cite{abrahams2001metallic,kravchenko2004metal,shashkin2005metal,spivak2010transport,shashkin2019recent}, which can exceed an order of magnitude. The phenomenon still lacks a comprehensive quantitative microscopic description.  Early theoretical work focused on the interplay between disorder and interactions using renormalization-group scaling theory \cite{finkelstein1983influence,finkelstein1984weak,castellani1984interaction,lee1985disordered,castellani1998metallic}; later, the theory was extended by Punnoose and Finkel'stein to take account of the existence of multiple valleys in the electron spectrum \cite{punnoose2001dilute,punnoose2005metal}.  This approach did allow for the existence of the metallic state, stabilized by the electron-electron interactions, in 2D systems, which is concurrent with experiments (see, \textit{e.g.}, Refs.~\cite{kravchenko1994possible,kravchenko1995scaling,popovic1997metal,coleridge1997metal,hanein1998the,shashkin2001metal,anissimova2007flow,melnikov2019quantum,melnikov2020metallic}).  According to this scenario, at temperatures well below the Fermi temperature, $T_{\text F}$, the resistivity $\rho$ should grow with the decreasing temperature reaching a maximum at $T=T_{\text{max}}$, and then decrease as $T\rightarrow0$.  The maximum in $\rho(T)$ dependence corresponds to the temperature at which the temperature-dependent screening of the disorder arises, and the interaction effects become strong enough to stabilize the metallic state and overcome the quantum localization.  This theoretical prediction, which is applicable only within the so-called diffusive regime (roughly, $k_{\text B}T\tau/\hbar<1$, where $\tau$ is the mean-free time), was found to be consistent with the experimental $\rho(T)$ data in silicon metal-oxide-semiconductor field-effect transistors (MOSFETs) \cite{punnoose2001dilute,anissimova2007flow,punnoose2010test}, but only in a narrow range of electron densities near the critical density $n_{\text c}$ for the metal-insulator transition.  In contrast, the corresponding strong changes in the resistivity with temperature are experimentally observed in a wide range of the electron densities: up to five times the critical density $n_{\text c}$, including the ballistic regime (roughly, $k_{\text B}T\tau/\hbar>1$), where the scaling theory is no longer applicable.\footnote{We stress that the ballistic regime introduced in Ref.~\cite{zala2001interaction} is not related to the well-known ballistic transport, or Knudsen regime, where the mean free path is larger than the sample dimensions.}

It should be noted, on the other hand, that according to Ref.~\cite{zala2001interaction}, a similar physical mechanism, namely, the elastic but temperature-dependent scattering of electrons by the self-consistent potential created by all other electrons (\textit{i.e.}, the Friedel oscillations), works in principle in both diffusive and ballistic regimes.  The interaction corrections in the corresponding limits are consistent with the logarithmic-in-$T$ corrections to the conductivity following from the renormalization-group scaling theory for diffusion modes
\cite{finkelstein1983influence,finkelstein1984weak,castellani1984interaction,lee1985disordered,castellani1998metallic,punnoose2001dilute,punnoose2005metal}, as well as with the linear-in-$T$ corrections to the conductivity predicted in earlier theories of temperature-dependent screening of the impurity potential \cite{stern1980calculated,gold1986temperature,dassarma1986theory,dassarma1999charged}, where the leading term has the form $\sigma(T)-\sigma(0)\propto T/T_{\text F}$; note that the Fermi temperature $T_{\text F}$ is in general determined by the effective electron mass $m$ renormalized by interactions.\footnote{The behaviors of the effective electron mass at the Fermi level and the energy-averaged effective electron mass are qualitatively different at low electron densities in the strongly correlated 2D system in SiGe/Si/SiGe quantum wells \cite{melnikov2017indication}, which is consistent with the interaction-induced band flattening at the Fermi level (see, \textit{e.g.},  Refs.~\cite{khodel1990superfluidity,nozieres1992properties,camjayi2008coulomb,zverev2012microscopic,bennemann2013novel,peotta2015superfluidity,volovik2015from,amusia2015theory}).  For the sake of simplicity, we will disregard this difference throughout this Rapid Communication.}  The theory of interaction corrections \cite{zala2001interaction} and the screening theory \cite{gold1986temperature} in its general form, which takes into account the mass renormalization, allowed one to extract the effective mass from the slope of the linear-in-$T$ correction to the conductivity in the ballistic regime \cite{shashkin2002sharp,shashkin2004comment}.  It was shown in Ref.~\cite{shashkin2002sharp} that the so-obtained effective mass sharply increases with decreasing electron density and that the $m(n_{\text s})$ dependence practically coincides with that obtained by alternative measurement methods \cite{shashkin2003spin,anissimova2006magnetization}.  However, corresponding small corrections calculated in the ballistic regime cannot convincingly explain the order of magnitude changes in the resistivity $\rho(T)$ observed in the experiment. In principle, in the spirit of the screening theories \cite{gold1986temperature,dassarma1999charged}, one can expect the resistivity to be a function of $T/T_{\text F}$ with a maximum at $T_{\text {max}}\sim T_{\text F}$, above which the electrons are not degenerate.  As of now, there are no accepted theoretical predictions allowing for a quantitative comparison with the experiment.

An alternative viewpoint in interpreting the temperature dependence of the resistivity is based on the so-called Wigner-Mott scenario, which focuses on the role of strong electron-electron interactions. The simplest theoretical approach to non-perturbatively tackle the interactions as the main driving force for the metal-insulator transition is based on dynamical mean-field theory (DMFT) methods of Refs.~\cite{camjayi2008coulomb,radonjic2012wigner,dobrosavljevic2017wigner} using the Hubbard model at half filling.  On the metallic side near the metal-insulator transition, the resistivity was predicted to initially increase as the temperature is reduced, reach a maximum, $\rho_{\text {max}}$, at temperature $T_{\text {max}}\sim T_{\text F}$, and then decrease as $T\rightarrow0$.  It was also shown that the resistivity change $\rho(T)-\rho(0)$, normalized by its maximum value, is a universal function of $T/T_{\text {max}}$.

Yet another approach to treat the strongly interacting 2D electron systems, focused on the Pomeranchuk effect expected within a phase coexistence region between the Wigner crystal and a Fermi liquid, was proposed in Refs.~\cite{spivak2003phase,spivak2004phases,spivak2006transport}. The predicted $\rho(T)$ dependence is also non-monotonic: the resistivity increases with decreasing temperature at $T\gtrsim T_{\text F}$ and decreases at lower temperatures.  However, no quantitative treatment of this problem, capable of quantitative comparison with experiment, currently exists.

\begin{figure}
\scalebox{0.45}{\includegraphics[angle=0]{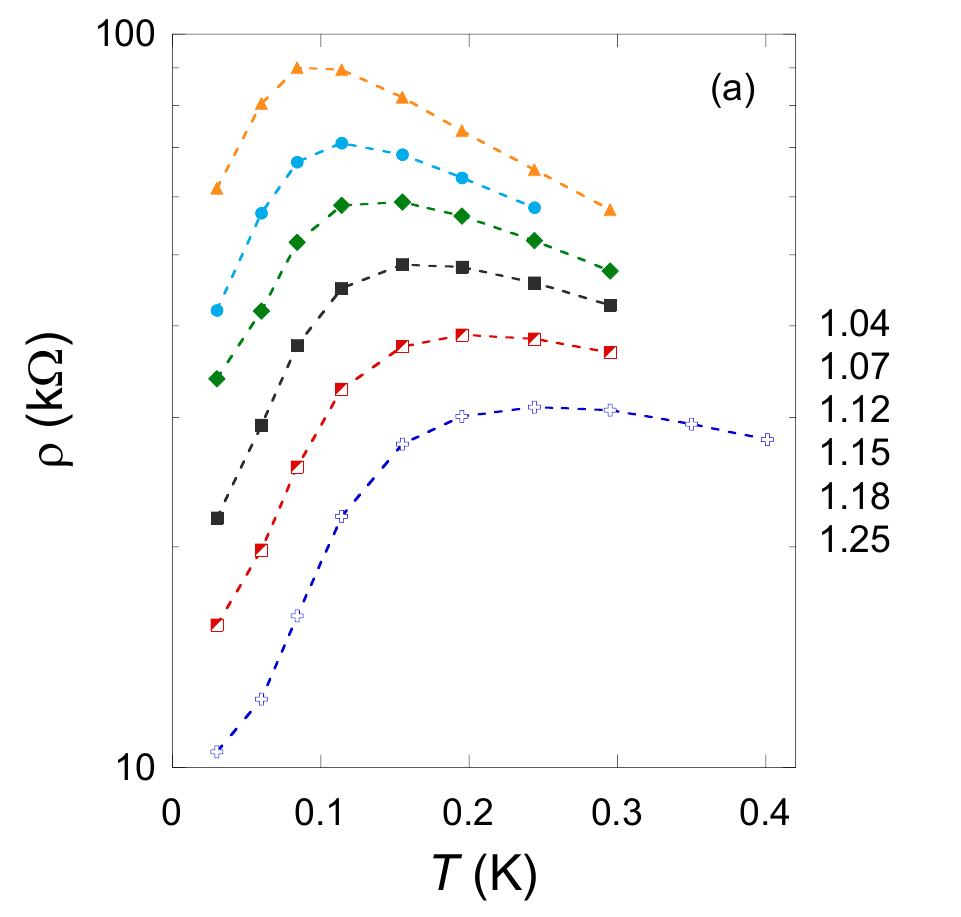}}
\scalebox{0.415}{\includegraphics[angle=0]{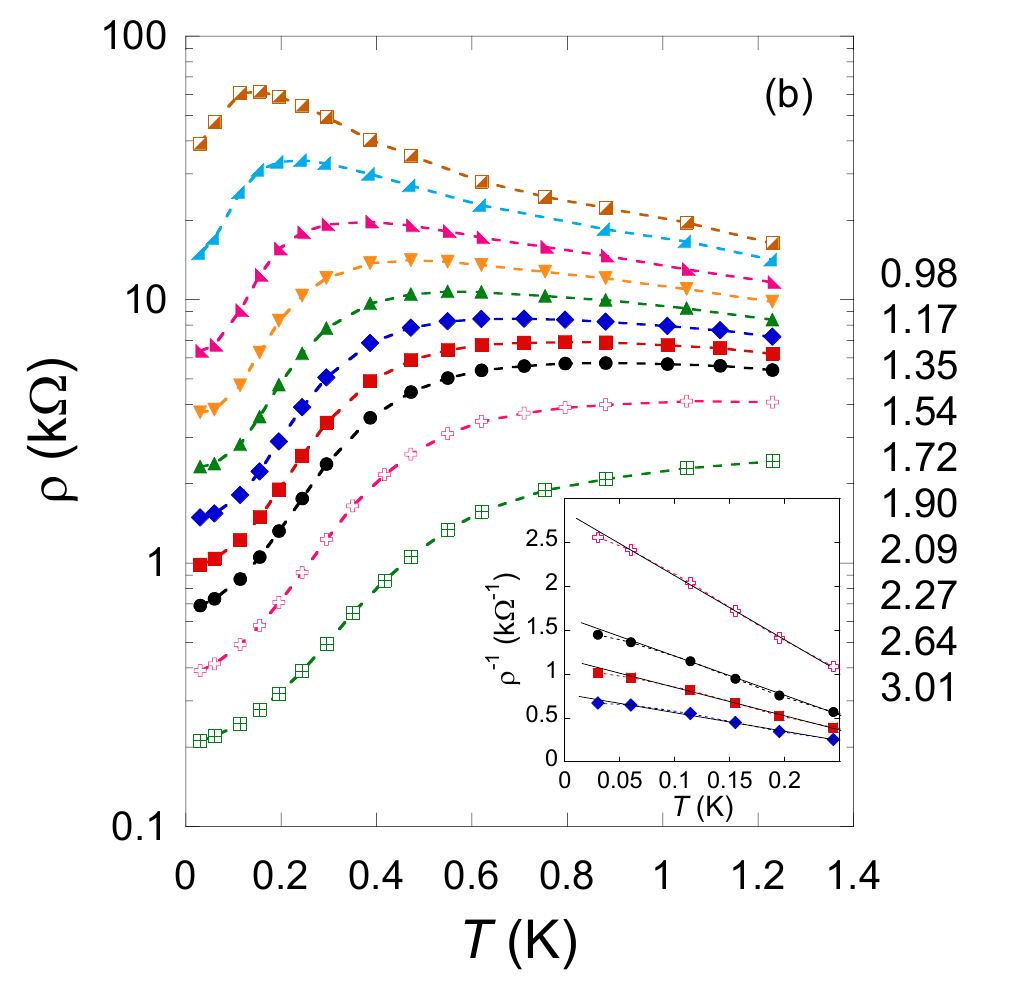}}
\caption{\label{fig1} Non-monotonic temperature dependences of the resistivity of the 2D electron system in SiGe/Si/SiGe quantum wells on the metallic side near the metal-insulator transition for samples A~(a) and B~(b).  The electron densities are indicated in units of $10^{10}$~cm$^{-2}$.  The inset in (b) shows $\rho^{-1}(T)$ dependences for four electron densities in sample B (the symbols are the same as in the main figure).  The solid lines are linear fits to the data.
}
\end{figure}

\begin{figure}
\scalebox{0.45}{\includegraphics[angle=0]{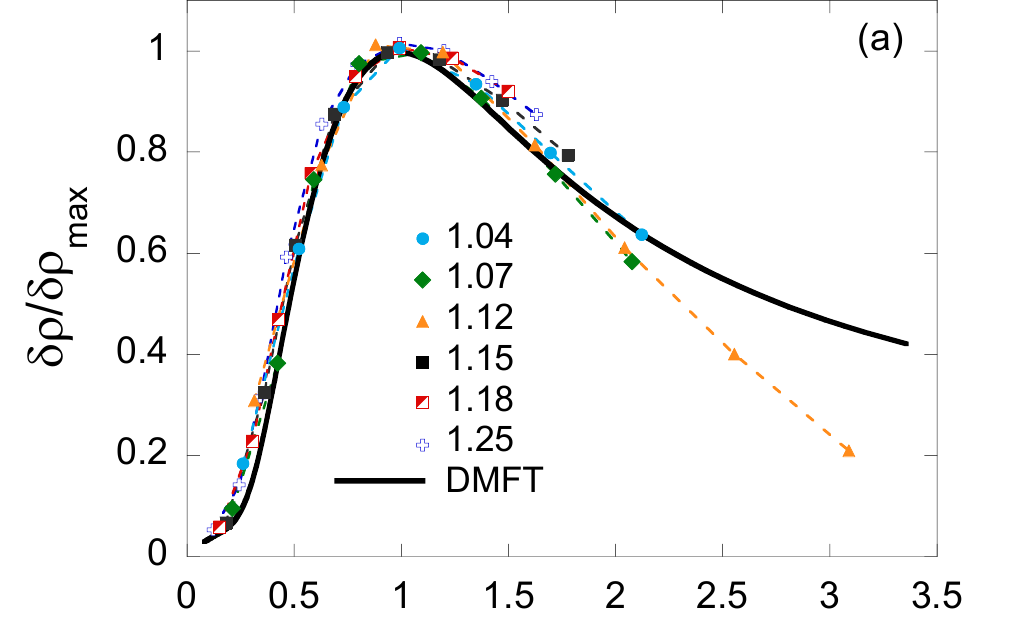}}
\scalebox{0.45}{\includegraphics[angle=0]{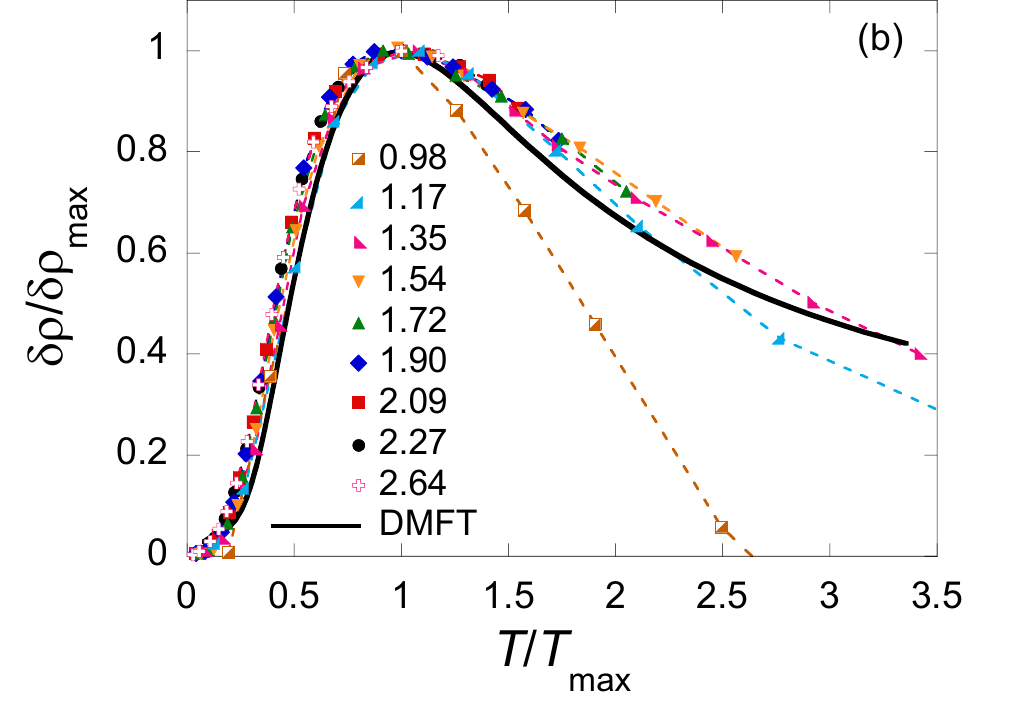}}
\caption{\label{fig2} The ratio $(\rho(T)-\rho(0))/(\rho_{\text {max}}-\rho(0))$ as a function of $T/T_{\text {max}}$ for samples A~(a) and B~(b).  Solid lines show the results of the dynamical mean-field theory in the weak-disorder limit \cite{camjayi2008coulomb,radonjic2012wigner,dobrosavljevic2017wigner}. The electron densities are indicated in units of $10^{10}$~cm$^{-2}$.
}
\end{figure}

\begin{figure}
\scalebox{0.459}{\includegraphics[angle=0]{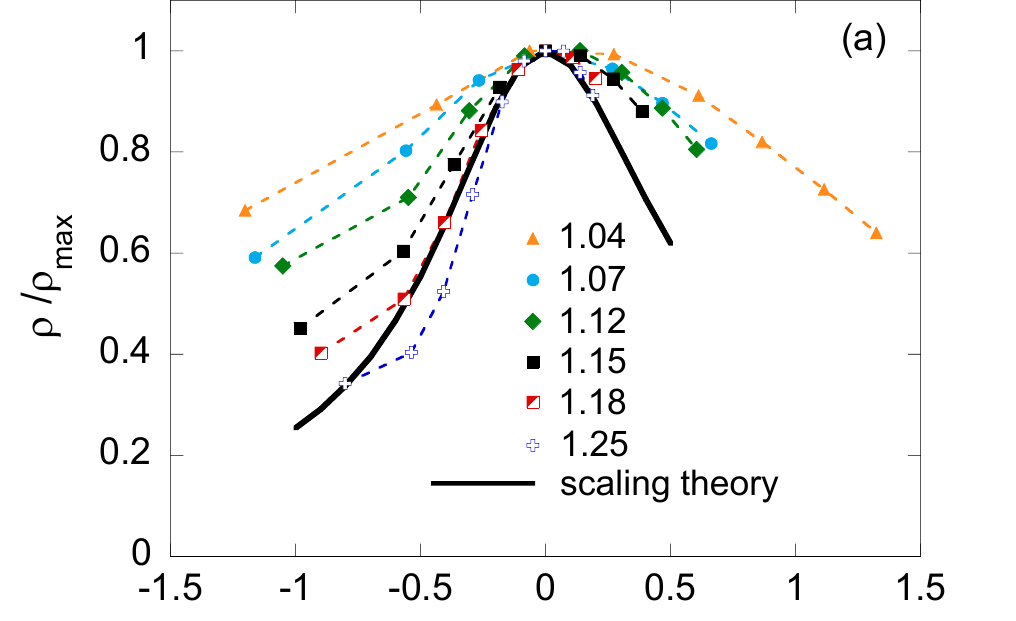}}
\scalebox{0.459}{\includegraphics[angle=0]{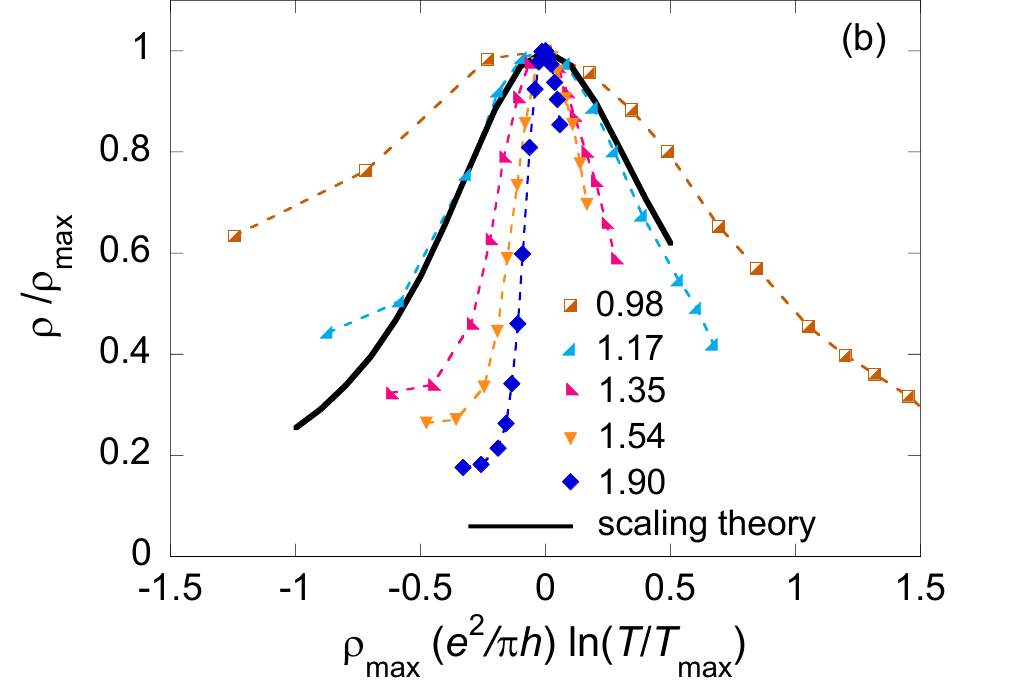}}
\caption{\label{fig3} The ratio $\rho/\rho_{\text {max}}$ as a function of the product $\rho_{\text {max}} \ln(T/T_{\text {max}})$ for samples A~(a) and B~(b).  Solid lines are the result of the renormalization-group scaling theory \cite{punnoose2001dilute,punnoose2005metal}.  The electron densities are indicated in units of $10^{10}$~cm$^{-2}$.
}
\end{figure}

To shed new light on the long-standing puzzle of the nature of the strong metallic temperature dependence of the resistivity in 2D electron systems, here we examine strongly correlated and ultra-clean SiGe/Si/SiGe quantum wells, in which the disorder potential is drastically weaker than that in the best silicon MOSFETs. We immediately observe that the resistivity, on the metallic side near the metal-insulator transition, increases with decreasing temperature, reaches a maximum at a temperature $T_{\text {max}}$, and then decreases by more than one order of magnitude.  The observed resistivity drop at $T<T_{\text {max}}$ in these samples is twice as large compared to the best 2D electron systems studied so far.  We scale our data in line with dynamical mean-field theory, according to which, the resistivity change $\rho(T)-\rho(0)$, normalized by its maximum value, is a universal function of $T/T_{\text {max}}$, and find a nearly perfect agreement with the predicted dependence in a wide range of electron densities except for the immediate vicinity of the metal-insulator transition, $(n_{\text s}-n_{\text c})\lesssim 0.1\ n_{\text c}$.  For comparison, we also perform the scaling analysis in the spirit of the renormalization-group scaling theory and find that, although the theory is consistent with the experimental results over a modest range of parameters, the data do not scale well in the wide range of the electron densities.  This is not particularly surprising because the scaling theory is expected to be valid only in the diffusive regime and at resistivity small compared to $\pi h/e^2$. Thus, the resistivity data are best described by the dynamical mean-field theory. Notably, similar behavior of the resistivity $\rho(T)$ can be expected within the screening theory in its general form, which adds confidence in both theories.

The samples studied are ultra-low disorder SiGe/Si/SiGe quantum wells similar to those described in detail in Refs.~\cite{melnikov2015ultra,melnikov2017unusual}. The peak electron mobility in these samples is 240~m$^2$/Vs, which is two orders of magnitude higher than that in the cleanest Si MOSFETs. The 15~nm wide silicon (001) quantum well is sandwiched between Si$_{0.8}$Ge$_{0.2}$ potential barriers. The samples were patterned in Hall-bar shapes with the distance between the potential probes of 150~$\mu$m and a width of 50~$\mu$m using standard photo-lithography. Measurements were carried out in an Oxford TLM-400 dilution refrigerator. The data were taken by a standard four-terminal lock-in technique in a frequency range 0.5--11~Hz in the linear regime of response.

Temperature dependences of the resistivity for two samples in the metallic regime are shown in Fig.~\ref{fig1} in the range of electron densities where the $\rho(T)$ curves are non-monotonic: at temperatures below a density-dependent temperature $T_{\text {max}}$, they exhibit metallic temperature behavior ($\mathrm{d}\rho/\mathrm{d}T>0$), while above $T_{\text {max}}$, their behavior is insulating ($\mathrm{d}\rho/\mathrm{d}T<0$). Note that the changes in the resistivity with temperature at $T<T_{\text {max}}$ are strong and may exceed an order of magnitude (more than a factor of 12 for the lowest curve in Fig.~\ref{fig1}(b)). The data recalculated into the conductivity as a function of temperature are displayed in the inset of Fig.~\ref{fig1}(b). Also shown are linear fits to the data. The observed linear temperature dependence is consistent with the ballistic regime not too close to the critical density $n_{\text c}$. As inferred from the temperature dependence of the conductivity, the transient region between ballistic and diffusive regimes corresponds to electron densities around $\approx 1.1\times10^{10}$~cm$^{-2}$.

The results of the scaling of our data for two samples in the spirit of dynamical mean-field theory \cite{camjayi2008coulomb,radonjic2012wigner,dobrosavljevic2017wigner} are shown in Fig.~\ref{fig2}.  The data scale perfectly in a wide range of electron densities and are described well by the theory in the weak-disorder limit; we emphasize that at some electron densities, the changes of the resistivity with temperature exceed an order of magnitude.  Deviations from the theoretical curve arise in the high-temperature limit in the transient region and become pronounced for $T>T_{\text {max}}$ at electron densities within $\sim10$\% of the critical value, which in these samples is close to $n_{\text c}\approx 0.88\times10^{10}$~cm$^{-2}$. The fact that in the low-temperature limit the same data display linear-in-$T$ corrections to the conductivity (see the inset in Fig.~\ref{fig1}(b)), which are in agreement with both the theory of interaction corrections \cite{zala2001interaction} and the generalized screening theory \cite{shashkin2004comment}, reveals the consistency of these theories and the DMFT. We argue that the DMFT can be applied to strongly interacting 2D electron systems.  Indeed, the Friedel oscillations near the impurities in real electron systems, even weakened by strong electron correlations \cite{andrade2010quantum}, signify that there is a short-range spatial charge order that plays the role of an effective lattice.  Note that the theory was also successful \cite{radonjic2012wigner,dobrosavljevic2017wigner} in quantitatively describing non-monotonic $\rho(T)$ dependences in silicon MOSFETs and $p$-GaAs heterostructures, although the changes in the resistivity were significantly weaker in those systems.

For proper perspective and comparison, we also perform a scaling analysis in the spirit of the renormalization-group scaling theory \cite{punnoose2001dilute,punnoose2005metal}, according to which the normalized resistivity $\rho/\rho_{\text {max}}$ should be a universal function of the product $\rho_{\text {max}}\ln(T/T_{\text {max}})$.  The results are plotted in Fig.~\ref{fig3}.  In both samples, only the data obtained at $n_{\text s}=1.18\times10^{10}$~cm$^{-2}$ for sample A (Fig.~\ref{fig3}(a)) and at $n_{\text s}=1.17\times10^{10}$~cm$^{-2}$ for sample B (Fig.~\ref{fig3}(b)) coincide nearly perfectly with the theoretical curve, although some deviations occur at the lowest temperature.  Pronounced deviations from the theory are evident at both higher and lower $n_{\text s}$.  At lower electron densities, the scaled experimental curves become wider than the theoretical one, and at higher densities, they become narrower.  A similar shrinkage of the scaled curves with increasing $n_{\text s}$ was reported earlier in Refs.~\cite{punnoose2001dilute,anissimova2007flow,radonjic2012wigner}.  One should take into account, however, that theory \cite{punnoose2001dilute,punnoose2005metal} has been developed for 2D electron systems that, on the one hand, are in the diffusive regime and, on the other hand, their resistivities are low compared to $\pi h/e^2$: at higher values of $\rho$, higher-order corrections become important and cause deviations from the universal scaling curve.  As a result, the applicable range of parameters becomes very narrow.

A question of how DMFT and the scaling theory are connected naturally arises.  Both theories predict non-monotonic temperature dependences of the resistivity.  Within the renormalization-group scaling theory \cite{punnoose2001dilute,punnoose2005metal}, the maximum in the $\rho(T)$ dependences occurs at a temperature well below $T_{\text F}$, at which the temperature-dependent interactions become strong enough to stabilize the metallic state and overcome the effect of the quantum localization.  This theory is relevant only in the diffusive regime.  Within the DMFT, in contrast, the maximum corresponds to the quasiparticle coherence temperature $T^\ast\sim T_{\text F}$: below this temperature, the elastic electron-electron scattering corresponds to coherent transport, while at higher temperatures the inelastic electron-electron scattering becomes strong and gives rise to a fully incoherent transport.  Even though the theoretical estimates of the positions of the maxima may be crude, the origins of the maxima are clearly different within these two theories in view of the role of the disorder. It should be stressed, on the other hand, that the functional forms of $\rho(T)$ dependences, including the maximum at $T_{\text {max}}\sim T_{\text F}$, expected from both the screening theory in its general form and DMFT, are similar. In particular, the linear temperature dependence of the conductivity at $T\ll T_{\text{F}}$ following from the generalized screening theory \cite{shashkin2004comment} and from the theory of the corrections to the conductivity due to the scattering on Friedel oscillations in the ballistic regime \cite{zala2001interaction} is consistent with the prediction of the DMFT.  The similarity of the theoretical predictions adds confidence in both theories and gives a hint that the underlying microscopic mechanism may be the same, \textit{i.e.}, electron-impurity or impurity-mediated electron-electron scattering for the strongly interacting case, as mentioned above.

Finally, we mention that similar non-monotonic $\rho(T)$ dependences are observed \cite{limelette2003mott,kurosaki2005mott} in quasi-two-dimensional organic charge-transfer salts (so-called Mott organics).  Interestingly, the DMFT is capable of quantitatively describing $\rho(T)$ dependences in these systems \cite{dobrosavljevic2017wigner}, which points out to the applicability of this theory to various strongly correlated systems.

Summarizing, we have observed that in a strongly interacting 2D electron system in ultra-low-disorder SiGe/Si/SiGe quantum wells, the resistivity on the metallic side near the metal-insulator transition increases with decreasing temperature, reaches a maximum at a temperature $T_{\text {max}}$, and then decreases by more than one order of magnitude. We have found that the normalized resistivity change $(\rho(T)-\rho(0))/(\rho_{\text {max}}-\rho(0))$ is a universal function of $T/T_{\text {max}}$ in a wide range of electron densities, which is in nearly perfect agreement with the dependence predicted by the dynamical mean-field theory. Notably, similar behavior of the resistivity $\rho(T)$ can be expected within the screening theory in its general form, which adds confidence in both theories. The renormalization-group scaling theory is found to be consistent with the experimental results within a modest range of electron densities near the metal-insulator transition, as expected.

A.A.S.\ and S.V.K.\ are grateful to A.~M.\ Finkel'stein for useful discussions. The ISSP group was supported by RFBR Grants No.\ 18-02-00368 and No.\ 19-02-00196 and a Russian Government contract. M.M.R.\ acknowledges the funding provided by the Institute of Physics Belgrade through the grant by the Ministry of Education, Science, and Technological Development of the Republic of Serbia.  Numerical simulations were run on the PARADOX supercomputing facility at the Scientific Computing Laboratory of the Institute of Physics Belgrade. The work in Florida was supported by NSF Grant No.\ 1822258 and the National High Magnetic Field Laboratory through the NSF Cooperative Agreement No.\ 1157490 and the State of Florida.  The NTU group acknowledges support by the Ministry of Science and Technology, Taiwan (Project No.\ 109-2634-F-009-029). The Northeastern group was supported by NSF Grant No.\ 1904051.

\end{document}